\documentclass[twocolumn,showpacs,preprintnumbers,prd,superscriptaddress,nofootinbib]{revtex4-1}

\usepackage{amsmath}
\usepackage{amsfonts}
\usepackage{amssymb}
\usepackage{graphicx}
\usepackage[colorlinks=true,linkcolor=blue,citecolor=teal,urlcolor=blue]{hyperref}
\usepackage{booktabs}
\usepackage{cleveref}
\usepackage{xcolor}
\usepackage{mathrsfs}
\usepackage{comment}

\usepackage[rightcaption]{sidecap}
\usepackage{subfigure}
\usepackage{soul}
\usepackage{enumitem}

\Crefname{equation}{Eq.}{Eqs.}
\Crefname{figure}{Fig.}{Figs.}
\Crefname{section}{Sec.}{Secs.}
\Crefrangeformat{equation}{Eqs.~(#3#1#4)--(#5#2#6)}

\begin{document}

\title{Geometric and topological corrections to Schwarzschild black hole}

\author{Rocco D'Agostino}
\email{rocco.dagostino@unina.it}
\affiliation{INAF - Osservatorio Astronomico di Roma, Via Frascati 33, 00040 Monte Porzio Catone, Italy}
\affiliation{Istituto Nazionale di Fisica Nucleare (INFN), Sezione di Napoli, Via Cinthia 21, 80126 Napoli, Italy}
\affiliation{Scuola Superiore Meridionale, Largo San Marcellino 10, 80138 Napoli, Italy}

\author{Orlando Luongo}
\email{orlando.luongo@unicam.it}
\affiliation{University of Camerino, Via Madonna delle Carceri, Camerino, 62032, Italy}
\affiliation{Department of Mathematics and Physics, SUNY Polytechnic Institute, Utica, NY 13502, USA}
\affiliation{INAF - Osservatorio Astronomico di Brera, Milano, Italy}
\affiliation{Istituto Nazionale di Fisica Nucleare (INFN), Sezione di Perugia, Perugia, 06123, Italy}
\affiliation{Al-Farabi Kazakh National University, Almaty, 050040, Kazakhstan}

\author{Stefano Mancini}
\email{stefano.mancini@unicam.it}
\affiliation{University of Camerino, Via Madonna delle Carceri, Camerino, 62032, Italy}
\affiliation{Istituto Nazionale di Fisica Nucleare (INFN), Sezione di Perugia, Perugia, 06123, Italy}

\begin{abstract}
In this paper, we compute departures in the black hole thermodynamics induced by either geometric or topological corrections to general relativity. Specifically, we analyze the spherically symmetric spacetime solutions of two modified gravity scenarios with Lagrangians $\mathcal{L}\sim R^{1+\epsilon}$ and $\mathcal{L}\sim R+\epsilon\, \mathcal{G}^2$, where $\mathcal{G}$ is the Euler density in four dimensions, while $ 0<\epsilon\ll 1$ measures the perturbation around the Hilbert-Einstein action.
Accordingly, we find the expressions of the Bekenstein-Hawking entropy by the Wald formula, and the black hole temperature and horizon of the obtained solutions. We then investigate the heat capacities in terms of the free parameters of the theories under study. In doing so, we show that healing the problem of negative heat capacities can be possible under particular choices of the free constants, albeit with limitations on the masses allowed for the black hole solutions.
\end{abstract}

\maketitle

\section{Introduction}

The debate concerning possible extensions of general relativity (GR) remains ongoing since cosmological observations have not definitively ruled out this possibility \cite{Clifton:2011jh,Joyce:2014kja,DAgostino:2018ngy,Berti:2015itd,Capozziello:2019cav,Ishak:2018his,DAgostino:2022tdk,Califano:2023aji,Muccino:2020gqt}. For example, the Starobinsky model of inflation \cite{Starobinsky:1980te}, emerging from a quadratic correction to the Hilbert-Einstein action, has proven in excellent agreement with the Planck data \cite{Planck:2018jri}\footnote{The Starobinsky model falls within the category of single large-field inflation models, which are susceptible to quantum gravity corrections at UV scales \cite{Ketov:2010qz,Kallosh:2014rga}. For alternative inflationary models, see e.g. \cite{DAgostino:2021vvv,DAgostino:2023tgm,Luongo:2023aaq,Belfiglio:2023rxb,Belfiglio:2024xqt,Belfiglio:2023moe,Belfiglio:2023eqi,Belfiglio:2022yvs}.}.
On the other hand, the dark energy problem plagues the standard cosmological picture based on GR \cite{Weinberg:1988cp,Carroll:2000fy,Peebles:2002gy,DAgostino:2019wko,Capozziello:2022jbw,Luongo:2021pjs}. Moreover, recent tensions between large-scale structures and the cosmic microwave background suggest that exploring theoretical scenarios beyond standard gravity is worth considering \cite{Riess:2021jrx,Abdalla:2022yfr,DAgostino:2020dhv,Perivolaropoulos:2021jda,DAgostino:2023cgx,Capozziello:2018hly}.

Although indirect probes cannot exclude departures from Einstein's model, a wide number of constraints severely limits the kind of alternatives to GR \cite{Ezquiaga:2017ekz,Amendola:2017orw,DAgostino:2019hvh,Bonilla:2019mbm,Luongo:2014qoa}. The latter, indeed, appears mostly favored in passing local-scale experiments \cite{Will:2014kxa,GRAVITY:2020gka}, while leaving the possibility open to corrections on the largest scales \cite{Koyama:2015vza}. However, the regime of strong gravity, where quantum gravity effects may naturally arise, has yet to be thoroughly explored. In this respect, the intense gravitational field near a black hole (BH) may hold the answer to revealing new physics and definitively confirming or challenging GR in favor of new hypotheses.

Since, thus far, no modified theory of gravity has been able to explain observations at all scales, it is conceivable that a gravitational action could include multiple contributions, each characterized by different significance at various scales. In this picture, stochastic processes occurring at a more fundamental level could lead to average quantities that appear as effective gravitational Lagrangians, showing only approximated behaviors in the limit of low energy. Therefore, in contrast to GR, one can consider higher than second-order field equations by including higher-order derivatives in the action, thus potentially providing possible solutions to the renormalizability problem at infrared scales \cite{Stelle:1977ry,Ketov:2022lhx,Bajardi:2023shq}. This is the case of $f(R)$ gravity action \cite{Sotiriou:2008rp,DeFelice:2010aj,Nojiri:2017ncd,Bajardi:2022ocw,DAgostino:2024sgm}, which assumes an arbitrary function of the Ricci scalar, $R$, or even more generic actions containing topological invariants, such as the Euler density in four dimensions, also known as the Gauss-Bonnet term, $\mathcal{G}$ \cite{Nojiri:2005jg,Li:2007jm,Bajardi:2022tzn}.

Bearing this in mind, we consider deviations in BH thermodynamics resulting from two distinct types of corrections to the Hilbert-Einstein action. Since different theories of gravity can admit the same BH solutions, although with distinct physical interpretations, we here distinguish between geometric and topological corrections, pinpointing a slight alteration in the geometry and an added second-order topological invariant within the action.
Recent studies in this direction have been undertaken, e.g., in Refs.~\cite{CamposDelgado:2022sgc,Delgado:2022pcc,Sajadi:2023bwe,Arora:2023ijd,Mustafa:2024zsx}, where BH solutions were investigated through a perturbative approach around the Schwarschild spacetime.
To assess the thermodynamic properties of the modified spacetime solutions, we utilize the Wald formula and derive the corresponding temperature by expanding around the primary term stemming from the Schwarzschild solution. This process allows us to quantify the primary deviations and deliberate on the feasibility of addressing the issue of thermodynamically stable BHs.

The present work is organized as follows. In Sec.~\ref{sezione2}, we introduce the geometric and topological corrections to the Hilbert-Einstein action to study spherically symmetric solutions. In Sec.~\ref{sezione3}, we characterize the thermodynamics of the obtained solutions, investigating potential regions in which one might find a possible resolution to the negative heat capacity issue typical of GR.
In Sec.~\ref{sec:conclusion}, we summarize our findings and provide the future perspectives of our work.

Throughout this paper, we consider the metric signature $(-,+,+,+)$, and set units of $c=\hbar=G=k_B=1$.

\section{Corrections to Schwarzschild spacetime}\label{sezione2}

Spherical symmetry represents the simplest approach to characterizing astrophysical objects. However,  configurations exhibiting such a symmetry are unstable from a pure thermodynamic perspective. Indeed, the problem of specific heat is still an open challenge of spherical symmetry in the context of GR as the specific heat appears negative if one employs the Schwarzschild metric as a benchmark spacetime \cite{Davies:1978zz}.
In this respect, Hawking solved the problem, by assuming that the net entropy might be done by the sum of two contributions, leading to the well-known \emph{information paradox} \cite{Hawking:1976de,Mathur:2009hf}.
However, we can wonder whether different types of corrections to the Hilbert-Einstein action can resolve the problem, without adding further terms to the entropy.

Two typologies of corrections are possible, i.e., geometric corrections related to extensions of Einstein's gravity and topological corrections in which topological invariants are plugged into the action of GR. For each of these two cases, we report below our findings on the modified spacetime geometries associated with them.

\subsection{Geometric correction to Schwarzschild BH}

To study geometric departures from Einstein's gravity, we focus our attention on the action \cite{Clifton:2005aj}
\begin{equation}
\mathcal{S}_g=\frac{1}{16\pi} \int \sqrt{-g}\,  \, R^{1+\epsilon}\,\,d^4x\,,
\label{azione_R}
\end{equation}
where $g$ is the determinant of the metric tensor. Expanding for small $\epsilon$, we have
$R^{1+\epsilon}= R+\epsilon R \ln R + \mathcal{O}(\epsilon^2)$, so that, the variational principle applied with respect to the metric tensor provides us with the field equations
\begin{align}
&R_{\mu\nu}(1+\epsilon)+\epsilon \ln R R_{\mu\nu}-\frac{R}{2}g_{\mu\nu}\left(1+\epsilon \ln R\right)\nonumber \\
&+\frac{\epsilon}{R^2}\big(\nabla_\mu R\, \nabla_\nu R-R\nabla_\mu \nabla_\nu R+R g_{\mu\nu}\Box R \nonumber \\
&-g_{\mu\nu}\nabla_\rho R\, \nabla^\rho R \big)+ \mathcal{O}(\epsilon^2)=0\,.
\end{align}

To solve the above field equations, we take into account the spacetime metric
\begin{equation}
ds^2=-F(r)dt^2+\frac{dr^2}{F(r)}+r^2d\Omega^2,
\label{metric}
\end{equation}
where $d\Omega^2\equiv d\theta^2+\sin^2 \theta\, d\phi^2$ and the lapse function reads
\begin{equation}
F(r)=1-\frac{2 M}{r}-\epsilon\, h(r)\,.
\label{eq:alpha}
\end{equation}
Here, the unknown function $h(r)$ accounts for the correction to the Schwarschild solution. In the following treatment, we consider $0<\epsilon\ll 1$ to study small deviations from GR, which is recovered for $\epsilon\rightarrow 0$.
The solutions of the field equations for the line element \eqref{metric}
are reported in \Cref{appendix}.
Specifically, after eliminating  $h^{(3)}(r)$ and $h^{(4)}(r)$ through \Cref{eq:FE22_R,eq:FE33_R},  and making use of \Cref{eq:FE11_R}, we find
\begin{equation}
r h''(r) (3 M-2 r)+h'(r) (6 M-2 r)+2 h(r)=0\,,
\end{equation}
whose general solution is given by
\begin{equation}
    h(r)=c_1(3M-r)+\frac{c_2}{r}\,,
\end{equation}
where $c_1$ and $c_2$ are arbitrary constants.
Hence, \Cref{eq:alpha} reads
\begin{equation}\label{deviazione_R}
F(r)=1-\dfrac{2M}{r}-\epsilon\left(3c_1 M-c_1 r+\dfrac{c_2}{r}\right)+\mathcal{O}(\epsilon^2)\,.
\end{equation}
Here, $c_2$ shifts the gravitational charge, namely the BH mass, and therefore can be arbitrarily fixed taking a given mass contribution. Nevertheless, $c_1$ modifies the functional behavior of the solution far from the singularity.

To better fix the free constants, we can investigate how departures induced on the horizon, $r_H$, occur as the lapse function vanishes. In particular, from $F(r)=0$ we obtain
\begin{equation}\label{horizon_R}
    r_H=r_S+\epsilon  \left(2 c_1 M^2+c_2\right)+\mathcal{O}(\epsilon^2)\,,
\end{equation}
where $r_S\equiv 2 M$ corresponds to the Schwarzschild radius. Since $c_2$ shifts the mass magnitude, it is possible to interpret it as a further mass contribution, bounding its sign to be positive and, hence, $2 c_1 M^2+c_2>0$ from \Cref{horizon_R}.

Now, we shall calculate the entropy according to the Wald formula \cite{Iyer:1994ys}
\begin{equation}
    S=-8\pi A\, \dfrac{\partial \mathcal{L}}{\partial R_{rtrt}}\bigg|_{r=r_H},
    \label{eq:Wald_entropy}
\end{equation}
where $A=4\pi r_H^2$ is the BH area.
Specifically, we write the Lagrangian density at the first order in $\epsilon$:
\begin{equation}
     \mathcal{L}=\frac{1}{16\pi}(R+\epsilon R \ln R)\,.
\end{equation}
Thus, using the results in Appendix \ref{appendix0} for the metric \eqref{metric}, we obtain
\begin{equation}
   \dfrac{\partial \mathcal{L}}{\partial R_{rtrt}}=-\frac{1}{32\pi}\left[1+\epsilon(1+\ln R)\right],
\end{equation}
where
\begin{equation}
    R=\frac{6 c_1(M-r) \epsilon  }{r^2}+\mathcal O\left(\epsilon ^2\right).
\end{equation}
We then arrive at
\begin{equation}
    S=\frac{A}{4}\left\{1+\epsilon\left[1+\ln \left(\frac{-3 c_1 \epsilon }{2 M}\right)\right]\right\} +\mathcal{O}\left(\epsilon ^2\right).
    \label{eq:entropy_R}
\end{equation}
The latter represents a correction to the Bekenstein-Hawking relation, whose deviation is $\sim 1+\ln \left(\frac{-3 c_1 \epsilon }{2 M}\right)>0$. From this, one can deduce that $c_1<0$.

\begin{figure*}
    \includegraphics[width=3.25in]{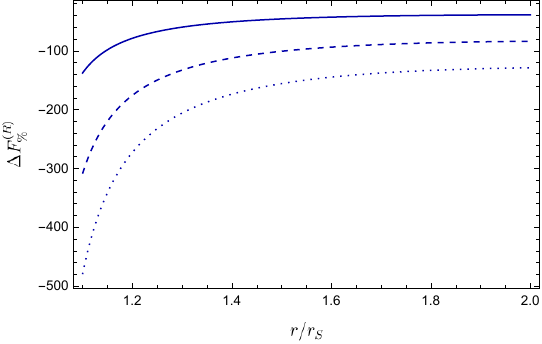}\qquad
    \includegraphics[width=3.35in]{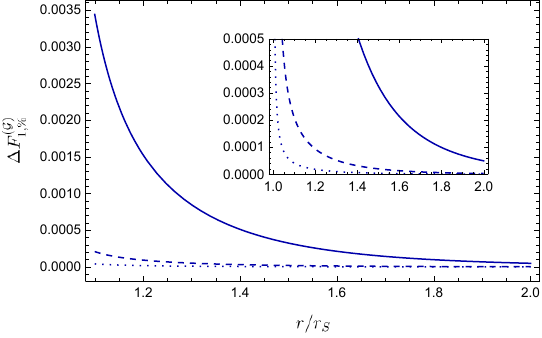}
    \caption{Relative percentage difference of the lapse function for the geometric  (left panel) and topological (right panel) corrections to GR, with respect to the Schwarschild solution. Conventionally, we choose $\epsilon=10^{-2}$, $c_1=-1$, $c_2=2M^2$, $k_1=k_2=0$. We display the cases for $M=5$ (solid line), $M=10$ (dashed line) and $M=15$ (dotted line).}
\label{fig:deltaF}
\end{figure*}

An indicative plot of the solution for the lapse function is displayed in \Cref{fig:deltaF} (left panel), where we show the quantity
\begin{equation}
    \Delta F_\%\equiv 100\left(\frac{F(r)-F_S(r)}{F_S(r)}\right)\,,
    \label{eq:Delta_F}
\end{equation}
with $F_S(r)=1-2M/r$ being the Schwarschild solution.
As an example, we can fix $c_1=-1$ and $c_2=2M^2$ corresponding to $r_H=r_S$, albeit $F(r)$ is different from the Schwarzschild lapse function\footnote{Notice that the horizon region for a generic spherically-symmetric spacetime can be the same even if the lapse functions are different for distinct metrics.}.

\subsection{Topological correction to Schwarschild BH}

As an example of topological correction to Einstein's gravity, we consider the  action
\begin{equation}
\mathcal{S}_g=\frac{1}{16\pi} \int d^4x\, \sqrt{-g}\,\left(R+\epsilon\, \mathcal{G}^2\right),
\label{azione_G}
\end{equation}
where \cite{Lovelock:1971yv,Fernandes:2022zrq}
\begin{equation}
    \mathcal{G}\equiv R^2-4R^{\mu\nu}R_{\mu\nu}+R^{\mu\nu\rho\sigma}R_{\mu\nu\rho\sigma}\,.
\end{equation}
The modified field equations, in this case, are given by
\begin{align}
&R_{\mu\nu}-\dfrac{1}{2}g_{\mu\nu}R+\bigg\{\frac{1}{2}g_{\mu\nu}\mathcal{G}^2+4\Big[(g_{\mu\nu}R-2R_{\mu\nu})\Box \mathcal{G} \nonumber \\
&+(-R\nabla_\mu\nabla_\nu+2R_\mu^\rho\nabla_\rho\nabla_\nu+2R_\nu^\rho\nabla_\rho\nabla_\mu)\mathcal{G}\nonumber \\
&-2(g_{\mu\nu}R_{\rho\sigma} +R_{\rho\mu\nu\sigma})\nabla^\sigma\nabla^\rho \mathcal{G}\Big]\bigg\}\epsilon+\mathcal{O}(\epsilon^2)=0\,.
\end{align}

Following similar previous works in the literature \cite{Calmet:2021lny,Xiao:2021zly,Delgado:2022pcc,Sajadi:2023bwe,CamposDelgado:2024jst}, we search for the solution of the field equations by considering the metric\footnote{Differently from the line element \eqref{metric}, a two-function metric is required to ensure that both the $tt$ and $rr$ components of the field equations are simultaneously satisfied for the model \eqref{azione_G}.}
\begin{equation}
    ds^2=-F_1(r)dt^2+\frac{dr^2}{F_2(r)}+r^2d\Omega^2,
\end{equation}
where
\begin{align}
    F_1(r)=1-\dfrac{2M}{r}-\epsilon\, h_1(r)\,, \quad F_2(r)=1-\dfrac{2M}{r}-\epsilon\, h_2(r)\,,
\end{align}
with $h_1(r)$ and $h_2(r)$ being two unknown functions to be determined from the field equations reported explicitly in \Cref{appendix}.
In particular, solving the \Cref{eq:FE11_G,eq:FE22_G} in the equatorial plane, we find
\begin{widetext}
\begin{align}
    h_1(r)= & -\frac{1}{385 M^5 r^{12}} \Big[ 645120\, M^{11}+16128\,  M^{10} r \left(55 r^2-42\right)-1408\, M^9 r^2 \left(720\,r^2-161\right)+12320\, M^8 r^3 \left(24 r^2+17\right)  \nonumber \\
    & +23760\, M^7 r^4+15840\, M^6 r^5+77\, M^5 r^6 \left(144-5 k_1 r^5\right)+8316\, M^4 r^7+6930\, M^3 r^8+6930\, M^2 r^9 \nonumber \\
    &+10395\, M r^{10}-10395\, r^{11}\Big]+\left(1-\frac{2 M}{r}\right)\left[\frac{27}{2 M^6}\ln \left(1-\frac{2 M}{r}\right)+k_2\right],
    \label{correz2_1}
\end{align}
and
\begin{align}
    h_2(r)& = -\frac{64\,M^3}{7 r^{10}}\left[63\, r \left(7+4  M^2\right)-288\, M r^2-938\,M+84\, r^3\right]+\frac{k_1}{r}\,,
    \label{correz2_2}
\end{align}
where $k_1$ and $k_2$ are integration constants.
\end{widetext}
In this case, requiring $F_1(r_H)=F_2(r_H)=0$, the horizon reads
\begin{equation}
    r_H= 2 M+\epsilon  \left(\frac{1}{M^5}-\frac{3}{7 M^3}+k_1\right)+\mathcal{O}(\epsilon^2)\,.
    \label{eq:horizon_GB}
\end{equation}

The behavior of the modified lapse function $F_1(r)$ is displayed in the right panel of \Cref{fig:deltaF}, in comparison with the standard case.

From the Lagrangian density of the theory, one obtains
\begin{equation}
    \dfrac{\partial \mathcal{L}}{\partial R_{rtrt}}=-\dfrac{1}{32\pi}\Big[1+2\epsilon\, \mathcal{G}(R+8g^{rr}R^{tt}-4R^{rtrt})\Big],
    \label{eq:der_L_GB}
\end{equation}
and using \Cref{correz2_1,correz2_2}, we find
\begin{equation}
    \dfrac{\partial \mathcal{L}}{\partial R_{rtrt}}=-\frac{1}{32\pi}\left(1+\frac{768  M^3 }{r^9}\epsilon\right)+\mathcal{O}\left(\epsilon ^2\right).
\end{equation}
In view of \Cref{eq:Wald_entropy,eq:horizon_GB}, we finally obtain
\begin{equation}
    S=\frac{A}{4}\bigg(1+\frac{3\epsilon }{2M^6}\bigg)+\mathcal{O}(\epsilon^2)\,.
    \label{eq:entropy_GB}
\end{equation}
This case is particularly interesting since the entropy is again corrected through a term $\sim 1+\frac{3\epsilon }{2 M^6}$ that, however, does not depend on the free integration constants, as obtained in the first case.

\section{Thermodynamics of modified gravity}\label{sezione3}

\begin{figure}
    \centering
    \includegraphics[width=3.3in]{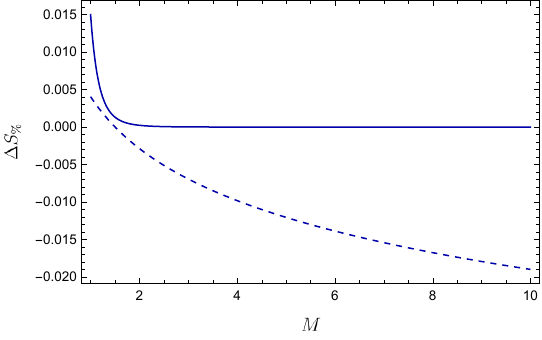}
    \caption{Relative percentage difference of the BH entropy as a function of the BH mass for the geometric (solid line) and topological (dashed line) corrections with respect to the GR prediction. We set $\epsilon=10^{-2}$ and $c_1=-1$.}
    \label{fig:entropy}
\end{figure}

We evaluate the corrections to GR entropy in Fig.~\ref{fig:entropy}, in which we show the quantity
\begin{equation}
    \Delta S_\%=100\left(\dfrac{S-S_H}{S_H}\right),
\end{equation}
where $S_H=1/A$ is the standard Hawking entropy, while $S$ is given by \Cref{eq:entropy_R,eq:entropy_GB} for the geometric and topological corrections, respectively.

Moreover, the corrections to the Hawking temperature can be computed as
\begin{equation}
    T_H=\frac{1}{4\pi}\frac{dF(r)}{dr}\bigg|_{r=r_H}\,.
    \label{eq:T_H}
\end{equation}
Specifically, from \Cref{deviazione_R,horizon_R} we obtain
\begin{align}
    T_H^{(R)}\simeq T_S-4\pi c_2 T_S^2\,\epsilon+\mathcal{O}(\epsilon^2)\,,
\end{align}
where $T_S=(8\pi M)^{-1}$ is the standard temperature induced by Hawking's entropy.

\begin{figure*}
    \includegraphics[width=3.3in]{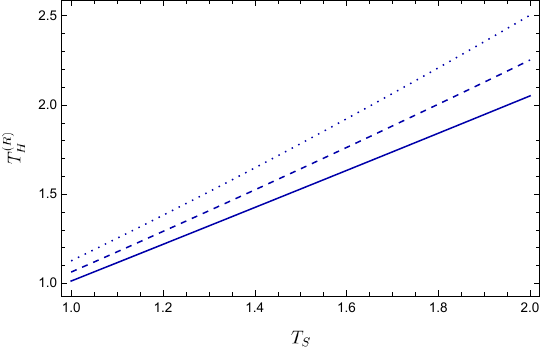}\ \qquad
    \includegraphics[width=3.3in]{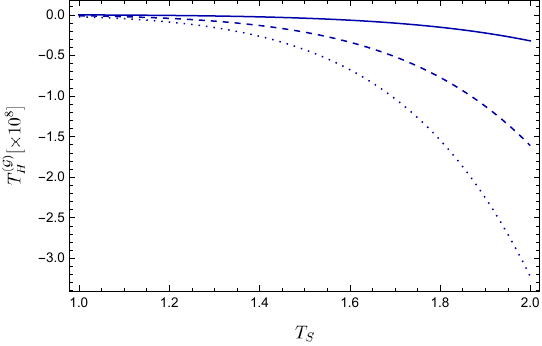}
    \caption{BH temperature as a function of Hawing's temperature for the geometric (left panel) and topological (right panel) corrections to GR. We display the cases for $\epsilon=10^{-3}$ (solid line), $\epsilon=5\times 10^{-3}$ (dashed line) and $\epsilon=10^{-2}$ (dotted line). The free constants are fixed as $c_2=-1$ and $k_1=-1/2$.}
    \label{fig:T}
\end{figure*}

For the topological correction, we instead find
\begin{align}
    T_H^{(\mathcal{G})}\simeq &\ T_S-\left(8 \pi  k_1 T_S^2-\frac{12288\, \pi ^4 T_S^5}{7}+262144\, \pi ^6 T_S^7\right)\epsilon \nonumber \\
    &+\mathcal{O}(\epsilon^2)\,.
    \label{temperatura2}
\end{align}
The behavior of the corrected temperatures is shown in \Cref{fig:T}, for different values of the $\epsilon$ parameter.
Interestingly, we observe that $T_H$ tends to decrease as $T_H^{(\mathcal{G})}$ increases, indicating that the larger the gravitational charge $M$, the lower the expected temperature. This result can be understood from \Cref{eq:entropy_GB}, as the dependence on the gravitational charge is inverse, scaling as $M^{-6}$ and dominating as the mass increases. This behavior suggests that one can investigate deviations from the Hawking temperature for supermassive BHs, making it particularly interesting to verify whether topological corrections to Einstein's gravity emerge in compact objects. In fact, the quite different results compared to the geometric correction case offer the opportunity to be experimentally tested in future experiments.

\subsection{Thermodynamic instability of heat capacities}

Information on the gravitational charge and thermodynamics of BHs is encoded in the relation
\begin{equation}\label{primoprincipio}
    dm=TdS\,,
\end{equation}
where $m$ is the effective BH mass, such that $m\neq M$ due to the correction to Einstein's gravity, or more broadly due to how much a BH is `hairy'.
In the Schwarzschild solution for GR, one has $M=m$,  recovered in our case as $\epsilon\rightarrow0$.

Then, one can compute the heat capacity at constant pressure as
\begin{equation}\label{standardC}
    \mathcal C=\frac{\partial m}{\partial T}=T\frac{\partial S}{\partial T}\,.
\end{equation}
In standard GR, as well-established, a \emph{negative heat capacity} is found, since $\mathcal{C}_S=-(8\pi  T_S^2)^{-1}$.
In our case, we can obtain the corrected mass by integrating \Cref{primoprincipio}:
\begin{align}
    m&=\int T(M)\frac{dS(M)}{dM}dM\,.
\end{align}
For the geometric and topological modifications of GR, we obtain
\begin{align}
    m^{(R)}&=M+\epsilon M\left[{\frac{3}{2}}  (1 + c_1 M)  +  \ln\left(\frac{ -3c_1}{2 M}\right)\right]  +\mathcal{O}(\epsilon^2)\,, \\
    m^{(\mathcal{G})}&= M+\epsilon\left(\frac{6}{5 M^5}-\frac{2}{7 M^3}-\frac{1}{2} k_1 \ln M\right)+\mathcal{O}(\epsilon^2)\,,
\end{align}
respectively.
Hence, \Cref{standardC} yields
\begin{align}
        \mathcal C^{(R)}=&-8\pi M^2 - 4 \pi \epsilon \bigg[ M^2 + 6 c_1 M^3 + 2c_2 M  \nonumber \\
        &+ 2 M^2 \ln\left(\frac{-3 c_1 \epsilon}{2 M}\right)\bigg]+\mathcal{O}(\epsilon^2)\,,
        \label{eq:C_R}
\end{align}
and
\begin{align}
        \mathcal C^{(\mathcal{G})}=&-8\pi M^2 -\epsilon\left(\frac{8 \pi }{M^4}-\frac{72 \pi }{7 M^2}+12 \pi k_1 M\right) +\mathcal{O}(\epsilon^2) \,.
        \label{eq:C_G}
\end{align}

We shall discuss in the following how, with an appropriate choice of constants, it is possible to change the signs of  $\mathcal C^{(R)}$ and $\mathcal C^{(\mathcal{G})}$, partially fixing the problem of negative heat capacity in modified theories of gravity.

\subsection{Physical consequences}

\begin{figure*}
    \includegraphics[width=3.1in]{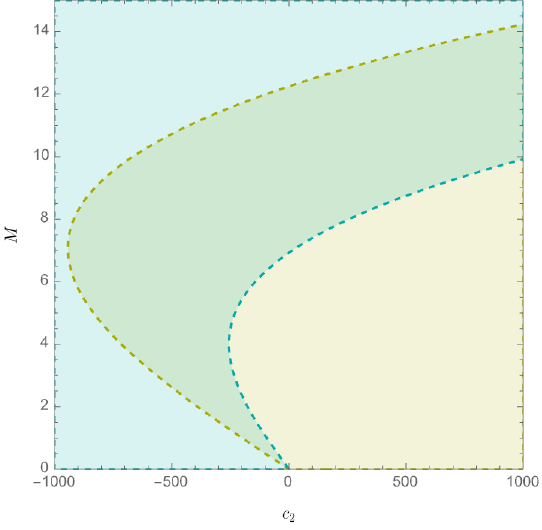}\hspace{1cm}
    \includegraphics[width=3.1in]{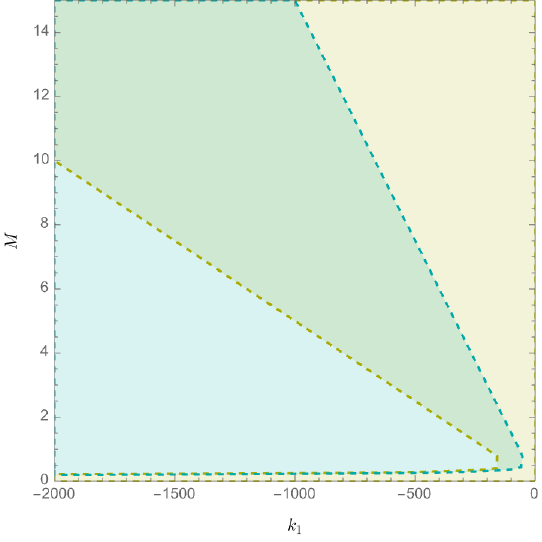}
    \caption{Parameter space for the BH solutions of the geometric (left panel) and topological (right panel) corrections to Einstein's gravity. We set $\epsilon=10^{-2}$ and $c_1=-2M/3$. The yellow and cyan regions correspond, respectively, to the conditions $r_H>0$ and $\mathcal{C}>0$.  The physically viable regions are only the green intersections, where both conditions are satisfied.}
    \label{fig:heat}
\end{figure*}

As we have shown, a term proportional to $r$ in the function $F(r)$ appears in our geometric correction to Einstein's gravity. Terms of such kind are present in conformal gravity \cite{Klemm:1998kf}, de Rham–Gabadadze–Tolley massive gravity \cite{deRham:2010kj} and, of course, in the $f(R)$ scenario \cite{Saffari:2007zt,Soroushfar:2015wqa}. In all the cases, the free parameters acquire very different physical meanings. In $f(R)$ gravity theories, the existence of a linear term in $F(r)$ can lead to modified Newtonian dynamics, i.e., to MOND theories \cite{Panpanich:2018cxo}, and then it appears suitable to work it out as a solution for small $\epsilon$. This term significantly affects the sign of specific heats, as we clarify below.

On the other hand, terms proportional to $r^{-n}$, with $n>1$, appear in our topological correction to GR.
These terms go to zero faster than multipole expansions of spherically symmetric spacetime, showing a non-trivial expansion of the lapse function.

Hence, whereas the geometric corrections are more sensitive at large distances, the topological ones behave instead differently. For the specific heats, we then expect quite different behaviors.
In particular, in the first model, $c_1$ and $c_2$ are not independent from each other. The heat capacity can change its sign for given sets of $c_1$ and $c_2$. In any case, the Minkowski limit is not fulfilled at large distances, where the term $\sim r$ dominates. Since $c_1<0$, $F(r)$ changes sign inevitably, indicating that the geometric modification is jeopardized by the issue of non-preserving the Lorentzian sign. This problem is not alone as the solution appears particularly degenerate due to the choice of the free constants that however cannot be chosen regardless of the underlying fine-tuning on the specific heat. Indeed, assuming large masses for the BH and setting conventionally $c_1=-r_S/3$, from Eq.~\eqref{eq:C_R} we obtain
\begin{equation}
    \mathcal{C}^{(R)}\simeq -8\pi M^2 -4\pi\epsilon\left(2c_2 M-4 M^4+2 M^2 \ln \epsilon +M^2\right).
\end{equation}
Thus, $\mathcal C^{(R)}\gtrsim0$ occurs for
   \begin{equation}
      c_2\lesssim 2 M^3-M \ln \epsilon -\frac{M}{2}-\frac{M}{\epsilon }\,.
      \label{cond1_c2}
   \end{equation}
Moreover, with the same choice of $c_1$, one might require (cf. Eq.~\eqref{horizon_R})
\begin{equation}
     r_H^{(R)}\simeq 2 M+ \epsilon  \left(c_2-\frac{4 M^3}{3}\right)>0\,,
\end{equation}
which provides us with the further condition
\begin{equation}
c_2\gtrsim 2M \left(\frac{2 M^2}{3}-\frac{1}{\epsilon }\right).
\label{cond2_c2}
\end{equation}

In the topological case, the heat capacity \eqref{eq:C_R} becomes positive if
\begin{equation}
    k_1\lesssim \frac{6}{7 M^3}-\frac{2}{3} \left(\frac{1}{M^5}+\frac{M}{\epsilon }\right),
\end{equation}
whereas, on the other hand, the positivity of the horizon radius \eqref{eq:horizon_GB} requires
\begin{equation}
    k_1\gtrsim \frac{3}{7 M^3}-\frac{1}{M^5}-\frac{2 M}{\epsilon }\,.
\end{equation}

In \Cref{fig:heat}, we display the BH mass ranges as functions of the free constants of the theories, such that both the BH horizon and the heat capacity are positive. Thus, the only viable solutions are those satisfying both conditions. We can observe that the parameter space allowing for a solution to the heat capacity problem is considerably wider in the case of the geometric correction to GR, compared to the topological correction. Nevertheless, it is worth remarking that the degree of arbitrariness in the choice of the free parameters limits the possibility of solving the aforementioned issue definitely.

\section{Final outlooks}
\label{sec:conclusion}

In this paper, we evaluated the main effects on BH thermodynamics induced by geometric and topological corrections to Einstein's gravity.
For this purpose, we first focused on a logarithmic correction to the Hilbert-Einstein action arising from a small power-law perturbation of the Ricci scalar, which is compatible with current experimental bounds on GR. As a second case, we analyzed a topological correction to Einstein's gravity obtained by adding the squared Gauss-Bonnet term in the gravitational action.
In both scenarios, we assumed that the terms responsible for the corrections cause a deviation from a spherically symmetric spacetime lapse function. In the first case, we showed that it is not possible to recover the Minkowski limit, whereas in the second case, the modified solution asymptotically goes to zero faster than the Schwarzschild one.

After inferring the modified event horizons, we investigated the corresponding BH thermodynamics through the Bekenstein-Hawking entropy by the Wald formula. We then analyzed the role of the arbitrary constants resulting from the solutions to the modified Einstein's field equations.
In particular, we computed the heat capacities, to check whether our corrections to Einstein's gravity can resolve the sign problem of GR.
Our results show that, for specific choices of the free constants, it is possible to obtain positive heat capacities within limited ranges of the BH mass. As it appears evident from Fig.~\ref{fig:heat}, the BH mass is intimately linked to the values of the free constants of the theories and as such, it is not possible to infer precise mass limits within which the heat capacity is positive. From a classical viewpoint, this fact remains a limitation of our approach since the constants cannot vary with respect to the BH mass.

Therefore, modifications of Einstein's gravity cannot fully resolve the issue connected to the specific heat sign, i.e., to the instabilities of spherical solutions. This may be seen as a consequence of the \emph{a priori} chosen actions. It is licit to presume that other classes of Lagrangians may allow obtaining a positive heat capacity in the whole mass ranges.

Future works will investigate if different gravity backgrounds may account for a definitive solution of the specific heat sign. Moreover, we will explore the role of alternative spacetimes, switching to cylindrical symmetry and possibly checking whether regular solutions exhibit analogous outcomes to our findings. Finally, it would be interesting to consider the running of the free constants of a given BH solution from a quantum perspective, where the issue of heat capacities could be healed through renormalization techniques.

\acknowledgments
The authors thank Ruben Campos Delgado for useful discussions and the referee for insightful comments on this work.

\begin{widetext}

\appendix

\section{Solutions of the field equations}
\label{appendix}

Here, we report the non-vanishing (diagonal) components of the field equations for the gravitational theories under consideration assuming the metric \Cref{metric}. Specifically, for the action \eqref{azione_R}, we have
\begin{align}
&\frac{2M-r}{2 r^3}\bigg\{\hspace{-0.1cm}-2h+r(2h'+rh''-r(4h'+rh''+\dfrac{2}{r(2h+r(4h'+rh'')^2} \Big[M(2h +r(4h'+rh'')(4h+2rh'-r^2(4h''+rh^{(3)})\nonumber \\
& +(2M-r)(-4h+r(-2h'+r(-2h'+r(-2h'+r(4h''(r)+rh^{(3)}))^2+(2h+r(4h'+rh'')(4(r-4M)h +r(4(M-r)h'\nonumber \\
&+r(2(2M+r)h''+r((6r-10M)h^{(3)}+r(r-2M)h^{(4)})))\Big]\bigg\}\epsilon +\mathcal{O}(\epsilon^2)=0\,,
\label{eq:FE11_R}
\end{align}
\begin{align}
&\left\{\frac{2rh^2+h(8r-12M+6r^2h'+r^3h''+r(4r^2h''^2
+h'^2+h'(4r-6M+r^3h''+r(3M-2r)(4h''+rh^{(3)})}{r^2 (2 M-r) \left(r \left(r h''+4 h'\right)+2 h\right)}\right\}\epsilon +\mathcal{O}(\epsilon^2)=0\,,
\label{eq:FE22_R}
\end{align}
\begin{align}
&\bigg\{\hspace{-0.1cm}-rh'-\dfrac{r^2}{2}h'+\frac{1}{r (r (r h''+4 h'+2 h))^2}\Big[(2M-r)(2h+r(4h'+rh'')(-4h+r(-2h'+r(4h''+rh^{(3)}))  \nonumber  \\
&+(2M-r)(-4h+r(-2h'+r(4h''+rh^{(3)}))^2 +(2h+r(4h'+rh'')(4(r-4M)h+r(4(M-r)h'+2r((2M+r)h'' \nonumber \\
&+r(3r-5M)h^{(3)}+r^3(r-2M)h^{(4)})\Big]\bigg\}\epsilon+\mathcal{O}(\epsilon^2)=0\,,
\label{eq:FE33_R}
\end{align}
\begin{align}
&\dfrac{ \sin^2\theta}{2}\bigg\{-2h+2(h+rh'-r(4h'+rh''+\frac{1}{r(2h+r(4h'+rh'')^2}\Big[-32Mh^2+4h((6r-32M)h' \nonumber \\
& +r(9r-16M)(2h''+rh^{(3)}+r^3(r-2M)h^{(4)}) +2r^2(4(2M-3r)h'^2 +r^2(2(22M-9r)h''^2+(2M-r)r^2{h^{(3)}}^2\nonumber \\
&+rh''((8M-3r)h^{(3)} +r(r-2M)h^{(4)}))+2rh'((8M+3r)h''+2r(2(3r-5M)h^{(3)}+r(r-2M)h^{(4)})))\Big]\bigg\}\epsilon+\mathcal{O}(\epsilon^2)=0\,,
\label{eq:FE44_R}
\end{align}
where $h^{(n)}$ denotes the $n$-order derivative of $h$ with respect to $r$.
\\

On the other hand, for the model \eqref{azione_G}, the diagonal components of the field equations read
\begin{align}
    &\frac{2M-r}{r^{13}}\Big\{1152 M^3 \left(4 r \left(3 M^2+7\right)-12 M r^2-r \cos (2 \theta ) (r-2 M)^2-67  M+3 r^3\right)-r^{10} \left(r h_2'+h_2\right)\Big\}\epsilon+\mathcal{O}(\epsilon^2)=0\,,
\label{eq:FE11_G}
\end{align}
\begin{align}
   &\frac{1}{r^{13} (r-2 M)^2}\Big\{(2 M-r) \left(r^{13}h_1'+1152  M^3 \left(4 M^2 \left(4-3 r^2\right) r-16 M^3+r^3 \cos (2 \theta ) (r-2 M)^2+ M \left(12 r^2-5\right) r^2-3 r^5\right)\right) \nonumber \\
   &+2 h_1  M r^{12}-h_2 r^{13}\Big\}\epsilon+\mathcal{O}(\epsilon^2)=0\,,
   \label{eq:FE22_G}
\end{align}
\begin{align}
  &-\frac{1}{2 r^{12} (r-2 M)^2}\Big\{(2M-r) \left(2304 M^3 (2 M-r) \left(8  M^3-8 M^2 r+4 r^3 \sin ^2\theta  (r-2 M)^2-33 M r^2+14 r^3\right)\right. \nonumber \\
   &\left.-r^{13} \left(r h_1'' (r-2 M)+h_1' (r-3 M)+h_2' (r-M)\right)\right)+2 h_1  M r^{12} (r- M)+2 h_2  M r^{12} (M-r)\Big\}\epsilon+\mathcal{O}(\epsilon^2)=0\,,
  \label{eq:FE33_G}
\end{align}
\begin{align}
  & \frac{\sin ^2\theta}{2} \bigg\{\frac{r (r-2 M) \left(r h_1'' (r-2 M)+h_1' (r-3 M)+h_2' (r-M)\right)+2 h_1  M (r-M)+2 h_2 M ( M-r)}{(r-2  M)^2}\nonumber \\
  &+\frac{2304 M^3 \left(8 M^3+8 M^2 r \left(2 r^2-1\right)- M r^2 \left(16 r^2+33\right)+2 r^3 \left(2 r^2+7\right)\right)}{r^{12}}\bigg\}\epsilon+\mathcal{O}(\epsilon^2)=0\, .
   \label{eq:FE44_G}
\end{align}

\section{Algebra of the Ricci scalar}
\label{appendix0}

The partial derivative of the Lagrangian in \Cref{eq:Wald_entropy} is computed by means of the following relations:
\begin{align}
    &\frac{\partial R}{\partial R_{\mu\nu\rho\sigma}}=\frac{1}{2}\left(g^{\mu\rho}g^{\nu\sigma}-g^{\mu\sigma}g^{\nu\rho}\right),\\
     &\frac{\partial (R_{\alpha\beta}R^{\alpha\beta})}{\partial R_{\mu\nu\rho\sigma}}=\dfrac{1}{2}\left(g^{\mu\rho}R^{\nu\sigma}-g^{\nu\rho}R^{\mu\sigma}-g^{\mu\sigma}R^{\nu\rho}+g^{\nu\sigma}R^{\mu\rho}\right),\\
      &\frac{\partial (R_{\alpha\beta\gamma\delta}R^{\alpha\beta\gamma\delta})}{\partial R_{\mu\nu\rho\sigma}}=2R^{\mu\nu\rho\sigma}\,.
\end{align}

\end{widetext}

\end{document}